\documentstyle[12pt]{article}
\begin{document}

\parskip 2mm plus 1mm \parindent=0pt
\def\cl{\centerline}
 \def\hs1{\hskip1mm}\def\hr{\hskip.2mm} 
\def\h10{\hskip10mm}\def\hsp{\hskip.5mm} \def\page{\vfill\eject}
\def\hx{\hskip10mm\hbox} \def\vs{\vskip4mm}
\def\<{\langle} \def\>{\rangle} \def\br{\bf\rm}  \def\it{\tenit}
 \def\de{\partial} \def\Tr {{\rm Tr}}\def\dag{^\dagger} 
\def\sadj{\hskip1mm{\rm sadj}}

\def\ne{=\hskip-3.3mm /\hskip3.3mm} 
\def\half{{\scriptstyle{1\over 2}}} \def\quter{{\scriptstyle{1\over 4}}} 
\def\Im{{\rm Im}} \def\Re{{\rm Re}} \def\I{{\rm I}} 
 \def\R{{\rm R}}  \def\M{{\br M}\hskip0.7mm} 
\def\G{{\bf G}}\def\F{{\bf F}}\def\H{{\bf H}}\def\L{{\bf L}}
 \def\P{{\rm P}} \def\Pr{{\br Pr}} \def\cc{{\rm cc}} \def\HC{{\rm
HC}} \def\D{{\br D}} \def\N{{\br N}} \def\S{{\rm S}}
\def\x{{\rm x}} \def\d{{\rm d}}\def\i{{\rm i}}\def\f{{\rm f}} 
\def\La{L_{\rm a}}\def\Lb{L_{\rm b}}\def\Lc{L_{\rm c}} 
 \def\Sl{S_{\rm loc}}
\def\Sc{S_{\rm c}} \def\ab{\overline a} 
\def\xv{\vec x}

\def\be{\begin{equation}}\def\ee{\end{equation}}

\vs\bf\cl{Quantum state diffusion, measurement and second quantization} 

\vs

\centerline {by}

\vs

\centerline {Ian C. Percival}\vs 

\centerline {Department of Physics} 
\centerline {Queen Mary and Westfield College, University of London} 
\centerline {Mile End Road, London E1 4NS, England}

\vskip15mm \centerline{\bf Abstract} \rm 

Realistic dynamical theories of measurement based on the diffusion of
quantum states are nonunitary, whereas quantum field theory and its
generalizations are unitary.  This problem in the quantum field
theory of quantum state diffusion (QSD) appears already in the
Lagrangian formulation of QSD as a classical equation of motion, where
Liouville's theorem does not apply to the usual field theory
formulation.  This problem is resolved here by doubling the number of
freedoms used to represent a quantum field.  The space of quantum
fields is then a classical configuration space, for which volume need
not be conserved, instead of the usual phase space, to which
Liouville's theorem applies.  The creation operator for the quantized
field satisfies the QSD equations, but the annihilation operator does
not satisfy the conjugate eqation.  It appears only in a formal role.

\vskip55mm
\vfill
 980615, QMW-PH-99-??\hfill Submitted to Phys. Lett. A
\page

%TODO

%??Amplify some of the explanations, which are too terse.

\vs\vs{\bf 1. Introduction}

Quantum measurement is a physical process by which the state of a
quantum system influences the value of a classical variable. The
meaning of {\it quantum measurement} here includes any such process,
including laboratory measurements, but also other, very different,
processes, such as the cosmic rays that produced small but detectable
dislocations in mineral crystals during the Jurassic era, and the
quantum fluctuations in the early universe that are believed to have
caused today's anisotropies in the universal background radiation and
in galactic clusters \cite{Percival1998c}.

Since Bohr and Einstein it has been recognized that it is difficult to
represent quantum measurement as a dynamical process
\cite{Wheeler1983a}.  Quantum theories that attempt it have problems
with unitarity.  These include those theories which depend on quantum
state diffusion (QSD), the principal subject of this letter
\cite{Percival1998c,Gisin1992c,Gisin1993d,Gisin1993b,Percival1994b}.

According to Bohr, the result of a measurement is influenced by the
conditions of the measurer.  Dynamical theories of quantum measurement
seek a dynamical process for this influence.  Quantum measurement
dynamics does not follow from the unitary dynamics of Schr\"odinger or
Heisenberg, nor from the quantum dynamics of fields, strings or
branes.  In a unified physics, they must be reconciled.

The methods of quantum field theory have been used for measurement
dynamics before, but this letter deals with the apparent contradiction
between the principles upon which they are based.  Quantum field
theory is unitary, whereas quantum state diffusion is not.  Further,
it has long been known that quantum measurement is nonunitarity
\cite{Luders1951,Gisin1984a}.

Here we trace the problem to the classical dynamics of a de Broglie
wave, considered as a classical field, in particular to the violation
of Liouville's theorem by the measurement process.  This makes it
necessary to reformulate the classical dynamics of the field
differently, making quantum measurement dynamics a nonlinear field
theory of a special type.

The experimental consequences of this theory are the same as the
standard results of quantum state diffusion when applied to the
dynamics of measurement, which is indistinguishable from the results
of the usual interpretation of nonrelativistic quantum theory for past
and current experiments, though not necessarily for all future
experiments \cite{Percival1998c}.  But the classical field theory of
quantum state diffusion has some unusual features, in particular that
the de Broglie wave is defined by a point in a configuration space,
not in a phase space.  There are canonical conjugate momenta, but
their role is probably formal, rather than physical, and they are not
the same as the complex conjugate amplitudes.  The reason for this
violation of one of the basic principles of field quantization is
given in the next section.

This letter is confined to the nonrelativistic formulation of the
dynamics of the single-particle de Broglie waves of QSD, first
as a classical field, then as a quantized field. 

\vs{\bf 2 Quantum state diffusion}

QSD represents measurement dynamics as a continuous stochastic
process, in which the state vector is the solution of an It\^o
stochastic differential equation \cite{Percival1998c}.  This is
expressed in terms of a complex differential stochastic fluctuation
$\d\xi$ with equal and independent fluctuations in its real and
imaginary parts, so that 
\be\label{a1} 
\M\d\xi = 0, \h10 \M(\d\xi)^2 =
0,\h10\M|\d\xi|^2 = \d t.  
\ee 
where $\M$ represents the mean over an ensemble.

Suppose a system with state $|\psi\>$ has Hamiltonian $\H$, and the
dynamical variable $G$ with Hermitean operator $\G$ is being measured.
According to QSD, measurement is a very rapid diffusion in state
space, whose rate is given by a real factor $c$ Then the quantum state
diffusion equation is 
\be\label{a2} 
{\d|\psi(t)\>\over\d t} =
-(i/\hbar)\H|\psi(t)\> -\half c^2\G_\Delta^2 |\psi(t)\> + c\G_\Delta
|\psi(t)\>{\d\xi\over\d t}, \ee 
where $\G_\Delta=\G-\<\psi(t)|\G|\psi(t)\>$ is the shifted
$\psi$-dependent $\G$-operator whose expectation for the current state
$|\psi(t)\>$ is zero.  For simplicity, we will absorb the constant $c$
into $\G$.  The equation is nonlinear, but the norm of $\psi(t)$ is
preserved.  The stochastic coefficient $\d\xi/\d t$ is a highly
singular function of time, whose singular properties are handled by
the It\^o calculus, but they need not concern us here.  What is
important is that it is a stochastic function of time representing
complex Gaussian white noise.

For laboratory experiments, the diffusion is so fast that the state
appears to jump between states on a time scale far shorter than the
other time scales of the system, in particular those of the
Hamiltonian.  However, according to quantum state diffusion theory,
this is a limiting case.  For small isolated quantum systems the
diffusion is so slow that it has not been detected.  This is
the other limiting case.

The details of quantum state diffusion as a theory of quantum
measurement are given in \cite{Percival1998c}. 

\vs{\bf 3 Classical measurement dynamics of the scalar field}

For simplicity, first consider a free particle with no measurement, in
a one-dimensional box with a bounded energy, so that there is a finite
number $N$ of discrete states.  In energy representation, with
energies $E_j = \hbar\omega_j$, the corresponding complex amplitudes
\be\label{a}
\psi_j= {1\over\sqrt{2}} (q_j+ip_j)
\ee
satisfy
\be \label{b}
i\dot \psi_j = \omega_j \psi_j,\hx{so that}
\ee\be \label{c}
\dot q_j = \{q_j,H\} = \omega_j p_j,\hskip4mm
\dot p_j = \{p_j,H\} = -\omega_j q_j,
\hskip4mm\hbox{with}\hskip4mm H 
= \sum_j{\omega_j\over 2}(p_j^2+q_j^2),
\ee
which are are Hamilton's equations for $N$ oscillators with real
canonically conjugate configuration and momentum coordinates
$q_j,p_j$.  Schr\"odinger evolution of the wave produces a unitary
transformation in the state space, a generalized rotation on the unit
sphere, identical to the motion of the phase point in the phase space
of the oscillators.  The unit sphere is an energy shell of a phase
space, Liouville's theorem is satisfied, so the phase space density
for a continuous distribution of systems is conserved.  Second
quantization of the field amplitudes follows just as first
quantization for the oscillators.

The same applies formally for the complex configuration coordinate
and its canonical conjugate momentum 
\be\label{d}
\psi_j={1\over\sqrt{2}}(q_j+ip_j)\hx{and}
\h10 i\psi_j^*={1\over\sqrt{2}}(iq_j+p_j).  
\ee
In this representation, the equations of motion for the configuration
and momentum coordinates are independent:
\be \label{e}
\dot\psi_j = -i\omega_j\psi_j,\hskip3mm
\dot\psi_j^* = i\omega_j\psi_j^*
\hskip6mm\hbox{with}\hskip6mm H = \sum_j\omega_j\psi_j^*\psi_j
=\sum_j(-i\omega_j)(i\psi_j^*)\psi_j.
\ee

The dynamics of quantum measurement is very different.  The norm of
the state is preserved, so it is confined to the unit sphere in state
space, but the density of states on the surface of the unit sphere is
not preserved.  Measurement of the energy, for example, puts the system
into one of the energy eigenstates, so a continuous distribution over
the unit sphere of an ensemble of systems is reduced towards a set of
at most $N$ points.  A uniform distribution over the unit sphere in
this finite-dimensional state space evolves towards a set of equal
$\delta$-distributions at each of the energy eigenstates.  The total
volume of the surface of the sphere is reduced towards zero.
Liouville's theorem is violated with a vengeance, so the space of
quantum states of the particle cannot be the phase space of any
classical Hamiltonian system.

However, the state space can be treated as a configuration space.
There is no conservation of volume in this configuration space, so a
Lagrangian or Hamiltonian measurement dynamics is possible.  The
equations of motion for the complex configuration coordinates of the
oscillators are independent of the equations for the conjugate
momenta, as they are when there is no measurement, but the conjugate
momenta are no longer the same as the complex conjugates of the
configuration coordinates.

\vs{\bf 4 Lagrangian theory of free-field QSD}

The configuration space trajectory for a time-independent dynamical
system with two configuration coordinates $q,q'$ is stationary for the
action integral
\be \label{g1}
S = \int_{t_0}^{t_1}\d t L(q,q')
\ee
Equivalent Lagrangians have action integrals that give the same
equations of motion.  

Before treating the QSD equations, consider a simpler classical model.
The equations of motion for a dynamical system with equivalent
Lagrangians
\be \label{g}
L = -q'\dot q + q'f(q),\h10 L' = q\dot q' + q'f(q)
\ee
are
\be \label{i}
\dot q = f(q),\h10  \dot q' = -q'\de f(q)/\de q
 \ee
and the momenta conjugate to $q,q'$ are 
\be \label{k}
p = \de L/\de\dot q = -q',\h10  p' = \de L/\de\dot q' = q.
\ee

We can identify $-q'$ and $p$, write them both as $p$, and similarly
we can write $p'$ as $q$.  The primed coordinates are then no longer
needed, as in the usual canonical theory of quantum fields.  But a
momentum then appears in the Lagrangian, as it does in quantum field
theory, but which is not normally allowed in classical dynamics.  The
quantum theory of a complex amplitude of a free linear field has just
this form with $f(q)=i\omega q$, where $q$ is complex, and $q'=q^*$,
its complex conjugate, which is treated as an independent canonical
coordinate, giving 
\be 
\label{k1} 
L = -q'\dot q + i\omega q q', \h10 L' = q\dot q' + i\omega q q'.  
\ee 
We can now identify $-q'$, $q^*$ and $p$ in the resultant Lagrange
equations.  This is consistent for this case because the equation of
motion for $q^*$ is just the complex conjugate of the equation of
motion for $q$.  Identifying $-q'$, $q^*$ and $p$ is a only a formal
problem for the theory of quantum fields and is very convenient in
practice..

However if $f(q)\neq cq$, the equation of motion for $-q'=q^*$ is not
the complex conjugate of the equation for $q$, so even if the
identification is made at some initial time, it will no longer hold
for later times.  This is what happens for the Lagrangian theory of
the wave $\psi$ in QSD.  For QSD we start with four independent
configuration coordinates $q,q',q^*,{q'}^*$.  The starred coordinates
$q^*,{q'}^*$ {\it are} complex conjugates of the coordinates $q,q'$,
but the primed coordinates are not conjugate momenta.  In QSD for a
Schr\"odinger field, the configuration coordinates corresponding to
$q,q^*$ are $\psi_j,\psi_j^*$.

The Lagrangian formulation of QSD for the measurement of a dynamical
variable $G$ of a particle follows from this approach.  We derive
the equations for the fields themselves, not the field components.  It
is convenient to express the total action, which is a function of the
configuration coordinates $\psi,\psi',\psi^*,{\psi^*}'$, as twice the
real part of a complex Lagrangian $\Lc$, which depends on all the
configuration coordinates except the last.  Consequently $\Lc^*$ is
independent of $\psi'$ and makes no contribution to Lagrange's
equation for $\psi$.

The action is
\be \label{m}
S = \int\d t (\Lc + \Lc^*),\hx{where}
\ee
\be  \label{n}
-i\Lc = -\int\d^3 x\hr\psi'\dot\psi-i\int\d^3 x\hr\psi'H\psi +
\int\d^3 x\hr\psi'Q\psi\hx{and}
\ee\be 
Q = Q(\psi^*\hr,\hr\psi)
= -\half\Big(G-\int\d^3 x\hr\psi^*G\psi\Big)^2 
+\Big(G-\int\d^3 x\hr\psi^*G\psi\Big)\d\xi/\d t.
\ee
The Lagrangian of equation (\ref{n}) is the form suitable for 
varying with respect to $\psi'$ and ${\psi^*}'$.  For
the variation with respect to $\psi$ and $\psi^*$, we
vary the equivalent Lagrangian obtained by partial integration. 

The variation of $S$ with respect to $\psi'$ is straightforward, the
$\Lc^*$ term does not contribute, and $\Lc$ gives the QSD equation for
$\psi$, and the derivative of $\Lc$ with respect to $\dot\psi$ gives the
definition of the momentum $p_\psi$.  Together they make Hamilton's
equations for $\psi$:
\be \label{o} 
\dot\psi = -iH\psi + Q\psi,\hskip5mm\hbox{(QSD),}
\h10 p_\psi = -i\psi'
\ee 
The corresponding operations with ${\psi^*}'$ and
$\dot\psi^*$ give Hamilton's equations for $\psi^*$ and ${\psi^*}'$
and as the action is real, the equations are just the complex
conjugates of those for $\psi$ and  $\psi'$.  
\be \label{o1}
\dot\psi^* = iH\psi^* + Q^*\psi^*,\hskip5mm\hbox{(QSD),}
\h10 p_{\psi^*} = i{\psi^*}'.
\ee
Since these are complex conjugate equations, $\psi^*$ remains
the complex conjugate wave for all time.

The equation for $\dot\psi'$ is given by the variation with respect to
$\psi$.  It is not nearly so simple as the QSD equation, as it
involves both the $\Lc^*$ and $\Lc$ terms.  The additional terms
ensure that even if initially $\psi'=\psi^*$, it does not remain so
for later times, as it does in the absence of measurement.  In this
way the (probably nonphysical) momentum space of $\psi',{\psi^*}'$ can
carry away the phase space volume that is lost by the motion of the
state in the configuration space of $\psi,\psi^*$.
\page
\vs{\bf 5 Second quantization}

Before going into the analysis, we note an essential difference
between this theory and the usual theory of second quantization.  In
both, the most important operator is the field creation operator
$\psi^*$.  

In the usual classical field theory, (we ignore the factors $i$), the
complex conjugate of the field amplitude $\psi^*$ is the canonical
conjugate of the amplitude $\psi$, and both are important for
quantization.  The operator $\psi$ is the field annihilation operator,
whose properties come from its commutation relations, and these in
turn are derived from the commutation relations for the conjugate
momentum.  $\psi^*$ is also the Hermitian conjugate operator of
$\psi$.  Both these roles are held by the same operator, which greatly
simplifies the theory.

For the quantization of the QSD equations, these roles are separated.
The annihilation operator corresponding to the creation operator
$\psi^*$ is the conjugate momentum operator ${\psi^*}'$, which is not
the same as the Hermitean conjugate operator $\psi$.  For convenience
we denote the annihilation operator by
\be\label{q}
{\psi^*}' = \psi^0.
\ee
The quantum state of a Schr\"odinger field with $n$ particles is given
by operating $n$ times on the vacuum with the creation operator
$\psi^*(t)$.  The Heisenberg equations of motion for $\psi^*(t)$ are
the same as Hamilton's equations (\ref{o1}) above.  But
because of the decoupling between the coordinate and momentum
equations, only the first of these equations is important.  As far as
we know, the second Heisenberg equation has no physical significance.
The annihilation operator $\psi^0(t)$ has an important formal role in
deriving Heisenberg equations, just as in ordinary quantum field
theory, but since only the creation operators are needed to obtain a
field from the vacuum, the annihilation operators appear to have no
other physical significance than this.

Because of the decoupling of the coordinate and momentum equations of
motion in the Heisenberg equations, the relatively complicated
Heisenberg equation for $\psi^0$ is not needed to get the physical
field.  However, without the momenta $\psi'$ and $\psi^0$, there would
be no Liouville theorem for the classical formulation of QSD, and
consequently no unitarity for the quantized field.  The price of
unitarity is additional fields that are not physical, as far as we
know.

\vs{\bf 6 Discussion}

There are two possible approaches to the quantum field theory of QSD.  In
the first, which is treated here, the QSD equations for the de Broglie
wave are derived from a classical Lagrangian, as a prelude to the
second quantization.  In the second, a quantum state diffusion term is
added to the second quantized equations of motion for the particle.

The physical difference between these two approaches, is that in the
first approach, the Liouville's equation is satisfied in the extended
phase space of the de Broglie wave, so the quantized theory can be
unitary.  This is consistent with the unitarity of standard quantum
field theory, but the price is the introduction of conjugate momenta
that appear to play no physical role.  In the second approach, any
diffusion terms will destroy the unitarity of the field theory, which
makes it very difficult or impossible to reconcile with 
the modern theory of fields, strings and branes.

\vskip20mm
\bibliography{~/icp}   
\bibliographystyle{abbrv}   %otherstyles% alpha, abbrv

\end{document}